# Free-standing Graphene by Scanning Transmission Electron Microscopy

*F. Song* [1,2], *Z.Y. Li* [1]*, *Z.W. Wang* [1], *L. He* [2], *M. Han* [2] and *G.H. Wang* [2]

[1] Nanoscale Physics Research Laboratory, School of Physics and Astronomy, University of Birmingham, Birmingham B15 2TT, UK

[2] National Laboratory of Solid State Microstructures, Nanjing University, Nanjing 210093, China

## Abstract

Free-standing graphene sheets have been imaged by scanning transmission electron microscopy (STEM). We show that the discrete numbers of graphene layers enable an accurate calibration of STEM intensity to be performed over an extended thickness and with single atomic layer sensitivity. We have applied this calibration to carbon nanoparticles with complex structure. This leads to the direct and accurate measurement of the electron mean free path. Here, we demonstrate potentials using graphene sheets as a novel mass standard in the STEM-based mass spectrometry.

---

* Corresponding author.  Z.Li@bham.ac.uk

# 1. Introduction

Scanning transmission electron microscopy (STEM) has become an indispensable tool in many fields of science community, including physics, materials, biology and nanotechnology [1-6]. For example, in structural biology, the ability of STEM in three-dimensional mass mapping of large protein assemblies often provides invaluable input in structural model building [7]. There are two approaches using STEM to obtain accurate mass mapping, absolute and relative measurements. The former involves an extensive calibration of the operational parameters of the microscope that is very difficult to carry out accurately and requires additional hardware to be fit into the microscope, hence the approach has only been taken up by a few dedicated groups worldwide until now [8-10]. A simpler approach is the measurement of electron scattering cross section using a mass standard [5]. A common mass standard used in structural biology is a hollow cylindrical shaped tobacco mosaic virus (TMV). They are about 300 nm in length and about 18 nm in diameter with an inner channel of ~4 nm. The problem for TMV mass standard is not only its poor accuracy, but also mass loss due to radiation damage or mass gain due to filling of central cavity [5]. These drawbacks could affect the precision of mass measurements. Recently, size-selected gold clusters have been applied to serve as a convenient mass standard for nanoparticles [11]. This has opened an alternative avenue for a quick and easy way of weighing nanoparticles on supports and gaining insight into their fine structure and shapes. In principle, this method could be extended to include all elements. However, practically this is difficult for light elements, since STEM intensity is proportional to $Z^a$, where Z is atomic number of the elements and $\alpha$ is in the range of 1.5-1.9 depending on the detector collection angle, sample thickness, and the Debye-Waller factor of the atomic species [1,12,13]. As a result, the light elements would have a much weaker image contrast than that of heavy elements.

In this study, we propose an alternative mass standard using free-standing graphene [14,15] sheets for light element mass measurements in STEM. We show that the discrete numbers of graphene layers enable an accurate calibration of STEM intensity to be performed. As an example of applications, we have applied the calibrated STEM intensity from graphene layers to gain insight of carbon nanoparticles of complex structure.

## 2. Materials and Methods

### *2.1 Preparation and transfer of graphene flakes*

To prepare thin graphene sheets, we used the established procedures of micromechanical patterning followed by repetitive cleavage of a highly-ordered-pyrolytic-graphite (HOPG) [14,16]. Patterned HOPG flakes were first pressed on a glass slide with a photo resistive spun of 0.5 µm thickness. After heating the sample at 110°C for 30 minutes, the visible flakes were removed. The remaining HOPG on the glass slide was peeled by a scotch tape for more than 40 times, and the remainders were then transferred to an acetone bath. The solution containing graphene flakes was drop casting on a holey Formvar filmed grid (XXBR customized TEM support). Once the acetone was evaporated, the graphene flakes would be held on the grid. To achieve a large density of the captured graphene flakes (10-100 pieces/cm$^{-2}$), the procedure was repeated over 10 times.

### *2.2 Carbon nanoparticles*

The carbon nanoparticles were prepared by plasma sputtering. In a radiofrequency sputtering chamber, 20 sccm of Ar was introduced with the final pressure of 18 Pa. The net input power was 400 W for a microcrystalline graphite target of 2 inches in diameter. The carbon particles were deposited on a Formvar filmed grid through a small nozzle, through which a differential pumping (20 Pa - 10$^{-3}$ Pa) was applied. The Ar pressure can be adjusted to control the morphology of carbon nanoparticles [17,18].

## 3. Results and Discussion

**Figure 1** displays a representative overview, by STEM imaging, of the graphene sheets, highlighted by light blue shade. The image was taken in a 200 KV Tecnai F20 STEM fitted with a high angle annular dark field (HAADF) detector. Here, large areas with uniformed STEM contrast strongly suggest the uniform thickness of graphene flakes prepared in this study. The variation of the intensity between the uniform areas gives clear indication that the graphene sheets in that region are folded or stacked together.

To establish direct correlation between the STEM intensity with the number of layers in graphene sheets, we apply an independent layer counting method by utilizing dark lines in bright field TEM images at the edge of graphene sheets. It has been demonstrated previously by several groups that edges and folding of a few layers (1, 2 and 4) of freely suspending graphene sheets are dominated by corresponding numbers of dark lines in TEM images [19-23]. The validity of this method has been cross-checked using Raman spectrum [19, 24, 25], nanobeam electron diffraction [21] and electron energy loss spectroscopy [22, 23]. To quantify the layer thickness of our graphene sheets, we searched over a large area of the sample for the freely suspending graphene sheets (for example, the area indicated by the arrows in Fig. 1). **Figure 2 (a, b)** displays two typical examples of such edges with thickness of 3 and 10 layers, respectively, in high-resolution bright field imaging. Care needs to be taken to discern Fresnel fringes from the graphene edges. One may see the image in Fig. 2(a) is in focus and presents rather poor contrast. This guarantees the accurate layer counting of the sample. On the contratry, the image in Fig.2(b) is a bit underfocus, leading to quite better contrast. Crossing of the two defocused fringes introduces the error of 1 layer. The STEM images taken from the corresponding edge area are shown in **Fig. 2 (c, d)**. The line profiles in **Fig. 2 (e, f)** across the step-edges (marked by dashed line in the STEM images) show clearly step-wise intensity variation between vacuum and the graphene sheets.

**Figure 3(a)** displays the HAADF intensity as a function of graphene layer number. A monotonic relationship is apparent. Here, to obtain the HAADF intensity values, we examined a relative clean area (with uniform contrast) over at least 100×100 pixels and we then plotted the histogram of the pixel intensity distribution for that particular layer thickness. An example for a 4 layers thick area is shown in inset of **Fig. 3(a)**. The mean value of the histogram and the standard deviation for each individual layer thickness is shown in **Figure 3**. The error bars here include contributions from both contamination and the background dark counts. It is inherently difficult to avoid contamination completely in obtaining HAADF-STEM images. The Birmingham Tecnai microscope is equipped with a dedicated cryo-capsule around the specimen area. This has been shown to reduce contamination building-up rate substantially. However, sometimes during prolonged scanning over the interested area by the focused electron beam, the gradual built-up contamination layer would leave a visible mark when zooming out of the scanned area. If this was the case, the data would be discounted and a new area would be studied. A close inspection of **Fig. 3 (a)** shows that HAADF-STEM intensity fits reasonably well into a linear function through the origin for graphene thickness up to 9 layers.

The resultant linear relation suggests that the multiple scattering is negligible in this ultra-thin film region. This would simplify greatly the subsequent data analysis procedures, as simple kinematical scattering models would be applicable [26]. To see how far the single scattering model can be extended to, we have purposely searched for thick graphene sheets to image. The results are presented in **Fig. 3(b)** together with that of thin layers shown in **Fig. 3(a)**. It is interesting to observe that the linear relation of the HAADF intensity persists to an exceptional large number of layers at around 50. This value extends the range that has been reported experimentally in the literature for the thickest graphene thin film (35 layers) [18, 22, 23], and is roughly 3 folds larger than the linear regime found in size-selected gold clusters in terms of atomic column depth (~15 atom height) in shell structure model [11]. The latter may be attributed to the weak electron scattering of the carbon atoms compared with gold.

The effect of camera length on the HAADF-STEM intensity has also been investigated. No intensity reversal was observed when the camera length changes between 150-520 mm, which corresponds to the detector's minimum inner acceptance angle, $\theta$, of 48-14 mrad. In the present study, the camera length of 285 mm ($\theta$ = 25 mrad) was used to ensure that most electrons detected are from incoherent thermal diffuse scattering while at the same time to maximize the signal to noise ratio. Using the single electron scattering approximation [26], the electron cross-section can be directly deduced from fitting the HAADF intensity-thickness relation with an exponential function, $I = A(1 - e^{-\zeta d})$, where $d$ is the layer thickness, and $A$ and $\zeta$ are fitting parameters. **Fig. 3(b)** shows a good fit over the entire investigated thickness up to 100 layers. Using the book value of graphite layer spacing of 0.34 nm [27], we obtain the cross section and the electron mean free path being $7.6(\pm 0.6) \times 10^{-5}$ nm$^2$ and $120 \pm 9$ nm, respectively. The values are in line with what have been reported in literature for amorphous carbon and diamond [28].

The HAADF intensity vs. thickness relation in **Fig. 3** can be served as a carbon mass calibration for HAADF-STEM imaging. The uncertainty of weighing the nanoobjects originates from the interaction volume between the beam and the mass standard. The Tecnai F20 machines provide the focused electron beam with the diameter of 0.4nm presently with the interaction volume of 42.7*Layers Å$^3$[11]. Considering the minimum counting uncertainty of 0.6 layers and the bulk graphite density [28], the atomic balance achieves the optimum precision of nearly 3 carbon atoms. Here we demonstrate this application in studying carbon nanoparticles with complex morphology. **Figure 4** shows a typical bright field TEM image (a) and STEM image (b) of the carbon nanoparticles produced by plasma sputtering. The images were taken using 200 KV Tecnai F20 at Nanjing University with the comparable setting-up as for the graphene study in Birmingham. These particles have overall spherical shape. However, STEM reveals high intensity at the centre, though energy-dispersive X-ray spectroscopy confirms that no foreign atoms are present within the core while electron energy loss spectroscopy (EELS) shows both the core and the shell have similar graphite-like chemical bonding (see supplementary materials for details). Nanobeam electron diffraction confirms that the particles are amorphous. To characterize these inhomogenous nanoparticles, we apply the calibration curve in **Fig. 3**. Here, the image in **Fig. 4 (b)** may be viewed as a quasi- 3D mass density mapping of the nanoparticles, shown in **Fig. 4(c)** in false colour. It is apparent that all of these particles have complicated core-shell structures, in which the core is formed by dense nuclear seeds that are surrounded by a porous shell. The densities of the core and shell can be obtained as $2.3 \pm 0.3$ g/cm$^3$ and $0.45 \pm 0.06$ g/cm$^3$, respectively. The former is close to the book value of graphite density, 2.09-2.23 g/cm$^3$ [29]. The revelation of this complex structure of carbon nanoparticles may provide foundation for developing applications for energy storage, such as enhancement of hydrogen adsorption (two order of magnitude higher) in vacuum condition as compared with that of graphite. The result will be reported in a separated publication.

## 4. Conclusions

In summary, we have demonstrated potentials using graphene sheets for mass standard for light element materials in STEM based mass spectrometry. By applying this mass standard to the carbon nanoparticles with complicated core-shell inhomogeneous morphology, we show that the core is graphite-like carbon whiles the shell has a porous structure. Previous attempts using conventional electron microscopy based elemental techniques such as EDX and EELS have failed in distinguishing such a core-shell structure of the carbon nanoparticles. The advance shown in the present study is made possible today partly due to our recent work in exploring potential in quantification STEM analysis [2, 11] and partly due to the recent surge of interests in graphene in the scientific community, which makes this material routinely available through various physical and chemical preparation routes. It is envisaged that the current proposal of using graphene for mass standard would be interesting not only for the materials scientists who work on light elements, but also potentially for biologists as an alternative to the commonly used tobacco virus as mass standard, in particularly for their structural investigations of low molecular weight biological materials. As compared to our previous progress using the coarse mass standards of carbon nanoclusters (in reference [12]), this alternative standards has many advantages. For example, they have nearly perfect 2D crystalline structure, their thickness can be well controlled and accurately measured. Therefore an improved data uncertainty can be expected. They are more robust. Potentially they can also replace the commonly used amorphous carbon thin film as the ultrathin and quantitative TEM/STEM supports.

## Acknowledgements

We thank Dr. Mi Yeon Song for her helpful assistance in preparing graphene samples. We acknowledge UK Engineering and Physical Sciences Research Council for supporting this project. The work in China is supported by the National Natural Science Foundation of China (Grant numbers: 90606002, 10674056, and 10775070) and the National Key Projects for Basic Research of China (Grant numbers: 2009CB930501, 2010CB923401). Supporting Information is available online from Wiley InterScience or from the author.

# Reference

[1] O.L. Krivanek, M.F. Chisholm, V. Nicolosi, T.J. Pennycook, G.J. Corbin, N. Dellby, M.F. Murfitt, C.S. Own, Z.S. Szilagyi, M.P. Oxley, S. T. Pantelides, S.J. Pennycook, Nature 464 (2010) 571.

[2] Z. Y. Li, N. P. Young, M. Di Vece, S. Palomba, R. E. Palmer, A. L. Bleloch, B. C. Curley, R. L. Johnston, J. Jiang, J. Yuan, Nature 451 (2008) 46.

[3] A. Y. Borisevich, A. R. Lupini, S. J. Pennycook, Proc. Natl. Acad. Sci. U.S.A. 103 (2006) 3044.

[4] P. R. Buseck, R. E. Dunin-Borkowski, B. Devouard, R. B. Frankel, M. R. McCartney, P. A. Midgley, W. M. Posfai, Proc. Natl. Acad. Sci. U.S.A. 98 (2001) 13490.

[5] S. A. Muller, A. Engel, Micron 32 (2001) 21.

[6] J. S. Wall and J. F. Hainfeld, Ann. Rev. Biophys. Biophys. Chem. 15 (1986) 355.

[7] A. K. Paravastu, R. D. Leapman, W.-M. Yau, R. Tycko, Proc. Natl. Acad. Sci. U.S.A. 105 (2008) 18349.

[8] A. Singhal, J. C. Yang, J. M. Gibson, Ultramicroscopy 67 (1997) 191.

[9] J. M. Lebeau, S. D. Finflay, L. J. Allen, S. Stemmer, Phys. Rev. Lett. 100 (2008) 206101.

[10] J. M. LeBeau, S. Stemmer, Ultramicroscopy 108 (2008) 1653.

[11] N. P. Young, Z. Y. Li, Y. Chen, S. Palomba, M. D. Vece, R. E. Palmer, Phys. Rev. Lett. 101 (2008) 246103.

[12] P. Hartel, H. Rose, C. Dinges, Ultramicroscopy 63 (1996) 93.

[13] Y. M. Zhu, H. Inada, L. Wu, J. Wall, D. Su, Hitachie M News 3 (2009) 2.

[14] K. S. Novoselov, Proc. Natl. Acad. Sci. U.S.A. 102 (2005) 10451.

[15] A. K. Geim, K. S. Novoselov, Nature Materials 6 (2007) 183.

[16] J. C. Meyer, C. O. Girit, M. F. Crommie, A. Zettl, Nature 454 (2008) 319.

[17] M. Han, C. Xu, D. Zhu, L. Yang, J. Zhang, Y. Chen, D. K., F. Song, G. Wang, Adv. Mater. 19 (2007) 2979.

[18] F. Q. Song, X. F. Wang, R. Powles, L. B. He, N. A. Marks, S. F. Zhao, J. G. Wan, Z. W. Liu, J. F. Zhou, S. P. Ringer, M. Han, G. H. Wang, Appl. Phys. Lett. 96 (2010) 033103.

[19] A. C. Ferrari, J. C. Meyer, V. Scardaci, C. Casiraghi, M. Lazzeri, F. Mauri, S. Piscanes, D. Jiang, K. S. Novoselov, S. Roth, A. K. Geim, Phys. Rev. Lett. 97 (2006) 187401.

[20] Z. Liu, K. Suenaga, P. J. F. Harris, S. Iijima, Phys. Rev. Lett. 102 (2009) 015501

[21] J. C. Meyer, A. K. Geim, M. I. Katsnelson, K. S. Novoselov, T. J. Booth and S. Roth, Nature 446 (2007) 60.

[22] T. Eberlein, U. Bangert, R. R. Nair, R. Jones, M. Gass, A. L. Bleloch, K. S. Novoselov, A. Geim and P. R. Briddon, Phys. Rev. B 77 (2008) 233406.

[23] M. Gass, U. Bangert, A. Bleloch, P. Wang, R. Nair, A. Geim, Nature Nanotech. 3 (2008) 676.

[24] Z. H. Ni, H. M. Wang, J. Kasim, H. M. Fan, T. Yu, Y. H. Wu, Y. P. Feng, Z. X. Shen, Nano Letters 7 (2007) 2758.

[25] D. Graf, F. Molitor, K. Ensslin, C. Stampfer, A. Jungen, C. Hierold, L. Wirtz, Nano Letters 7 (2007) 238.

[26] I. Angert, C. Burmester, C. Dinges, H. Rose and R. R. Schroder, Ultramicroscopy 63 (1996) 181.

[27] A. B. Yen, B. E. Schwickert, Appl. Phys. Lett. 84 (2004) 4702.

[28] K. Iakoubovskii, K. Mitsuishi, Phys. Rev. B 77 (2008) 104102.

[29] H. H. Hsieh, Y. K. Chang, W. F. Pong, M.-H. Tsai, F. Z. Chien, P. K. Tseng, I. N. Lin, H. F. Cheng, Appl. Phys. Lett. 75 (1999) 2229.

*Figure Captions*

**Figure 1** HAADF-STEM image showing an overview of graphene flakes supported by holey Formvar film covered Cu grids. The graphene are highlighted by blue shades. The bright patches on the right side of the images are contamination left on the samples during preparation and handling of the specimen. The arrows indicate areas where the graphene freely are suspended on the holey film.

**Figure 2. Graphene thickness analysis.** (a, b) High-resolution TEM images of the folded graphene edge, where the dark lines indicate the number of layers of 3 and 10 for two cases. (c, d) The HAADF-STEM images of the corresponding areas. (e, f) The intensity line profiles averaged over a few pixels, taken from the positions indicated in (c) and (d).

**Figure 3. The calibration curve.** (a) The HAADF intensities are plotted against the layer number up to 9. The data and the error bars are taken from the mean value and the corresponding standard deviation from the selected layer numbers of graphene. The linear fitting is a guide for eyes. The inset shows an intensity histogram for a suspended 4-layer graphene. (b) An extended range of intensity-thickness relation is fitted by the single scattering approximation (solid line): $I = A(1 - e^{-\xi d})$. The linear fitting in (a) is shown here in dashed line as a comparison.

**Figure 4. Complex inhomogenous structure of carbon nanoparticles.** (a) TEM image of carbon nanoparticles prepared by plasma sputtering. (b) HAADF-STEM image of the carbon particles (not the same area as in (a) but from the same specimen). (c) The 3D view of mass mapping of nanoparticles

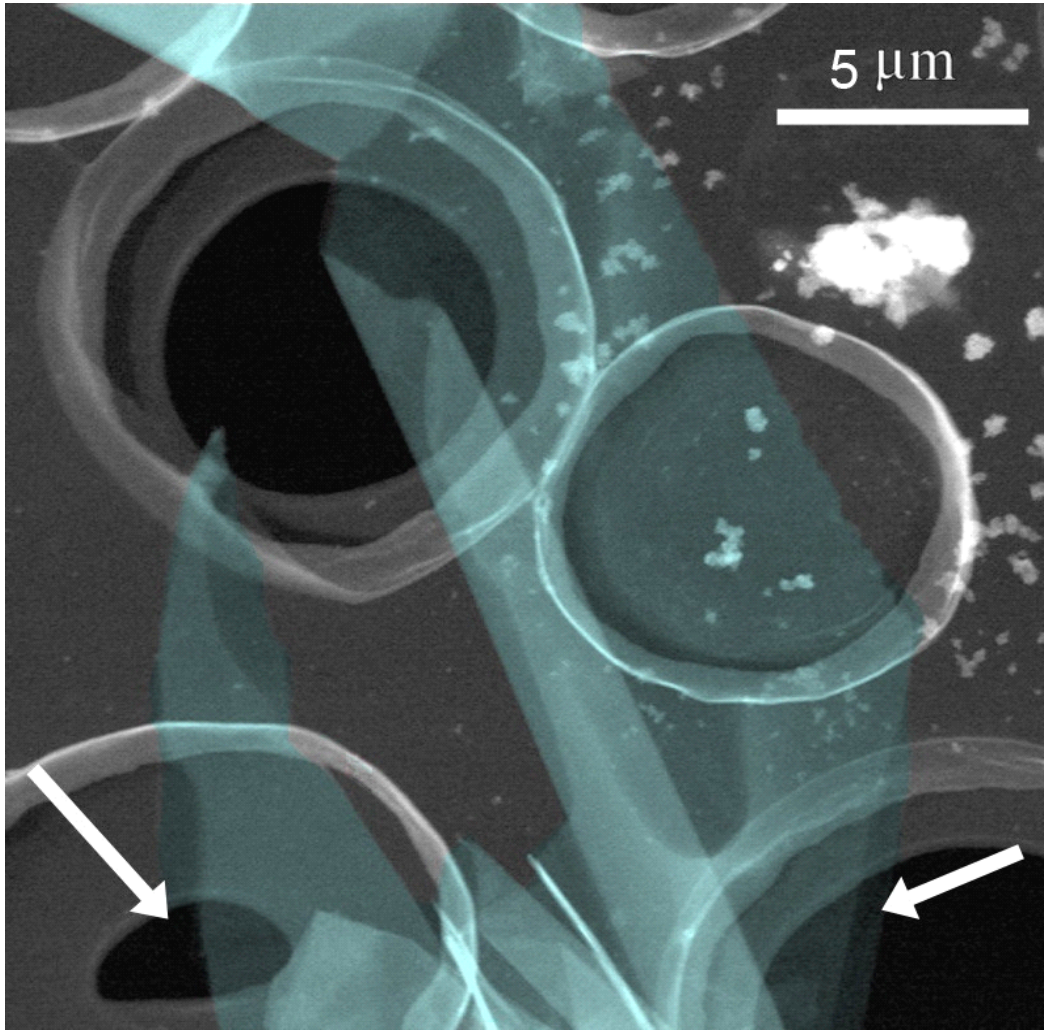

**Song et al. Figure 1**

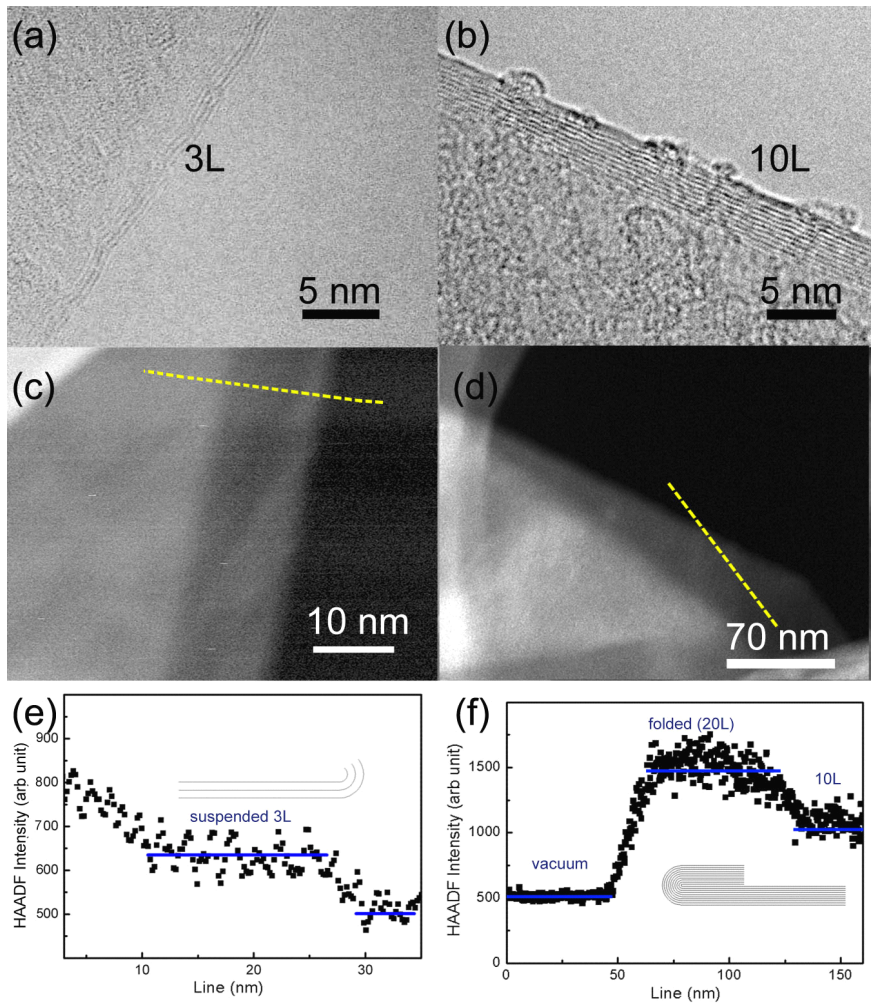

Song et al. Figure 2.

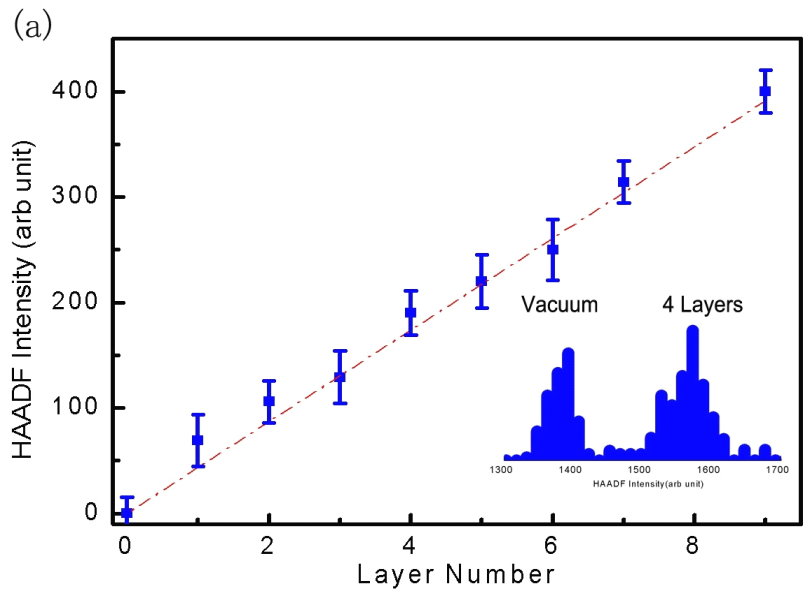

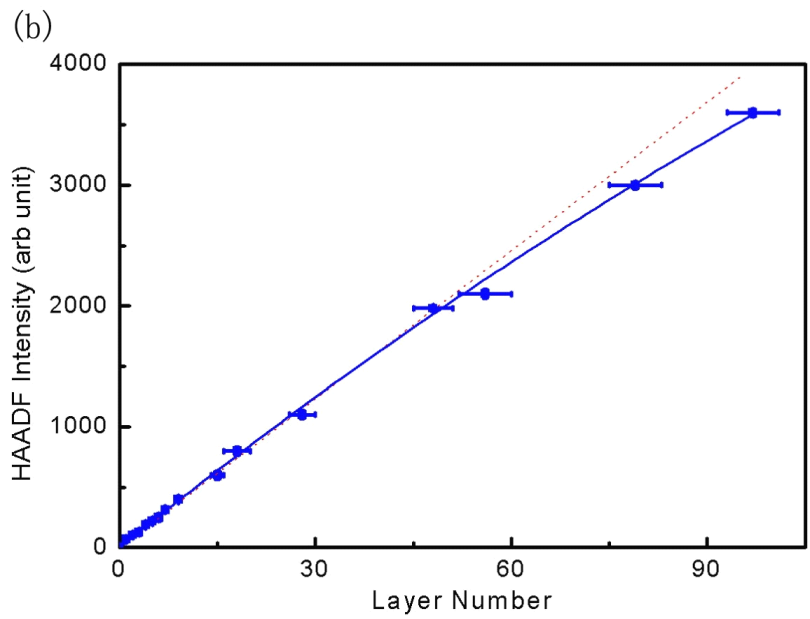

**Song et al. Figure 3.**

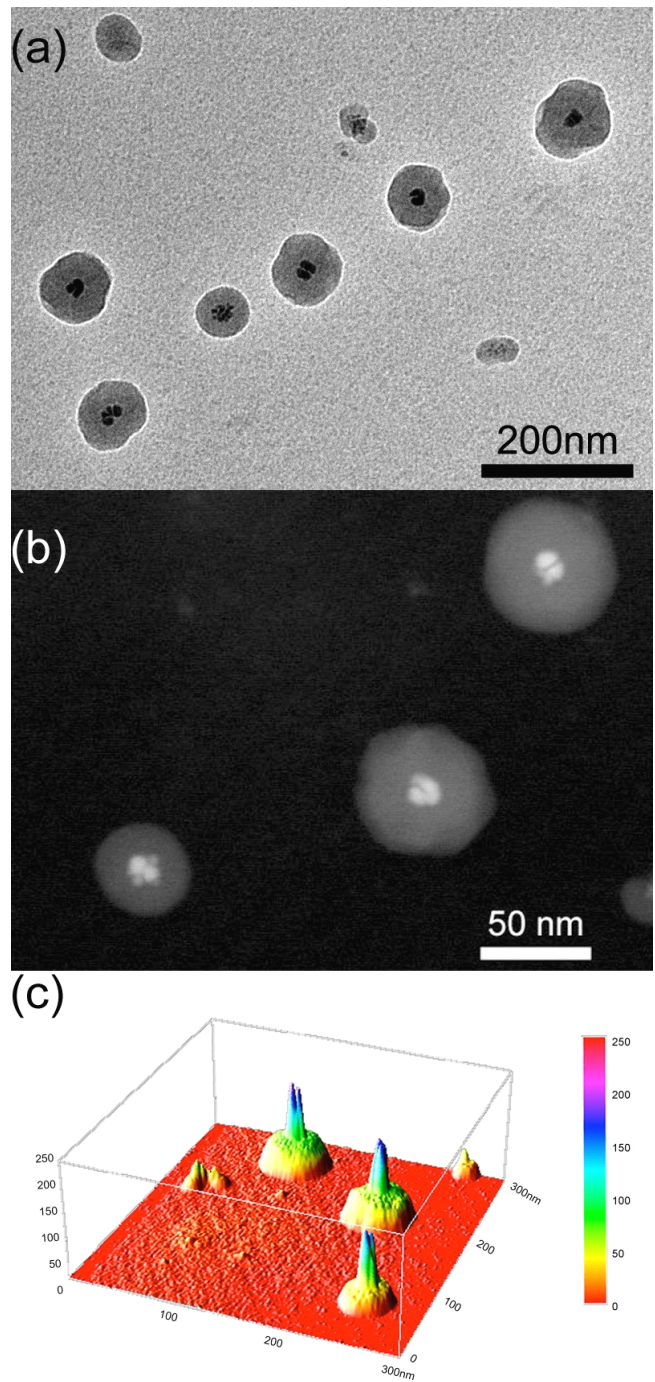

**Song et al. Figure 4.**